\documentclass[english,twocolumn,showpacs,preprintnumbers,amsmath,amssymb]{emulateapj}
\usepackage[T1]{fontenc}
\usepackage[latin9]{inputenc}
\usepackage{graphicx}
\usepackage{amssymb}

\makeatletter

\def\mn{\rm min} 
\def\mx{\rm max} 

\makeatother

\usepackage{babel}

\begin{document}

\title{{\normalsize Kozai cycles, tidal friction and the dynamical evolution
of binary minor planets}}

\author{Hagai B. Perets\altaffilmark{1} and Smadar Naoz\altaffilmark{2}}

\email{hagai.perets@weizmann.ac.il}

\altaffiltext{1}{ Weizmann Institute of Science, POB 26, Rehovot
76100, Israel} \altaffiltext{2}{Raymond and Beverly Sackler School
of Physics and Astronomy, Tel Aviv University, Tel Aviv 69978, Israel}
\begin{abstract}
In recent years many binary minor planets (BMPs) have been discovered
in the Solar system. Many models have been suggested for their formation,
but these encounter difficulties explaining their observed characteristics.
Here we show that secular perturbations by the Sun (Kozai mechanism)
fundamentally change the evolution and the initial distribution of
BMPs predicted by such models and lead to unique observational signatures.
The Kozai mechanism can lead to a large periodic oscillations in the
eccentricity and inclination of highly inclined BMP orbits, where
we predict such effects to be observable with current accuracy within
a few years (e.g. for the binary asteroid Huenna). In addition, the
combined effects of the Kozai mechanism and tidal friction (KCTF)
drives BMPs into short period circular orbits. We predict a specific
inclination dependent distribution of the separation and eccentricity
of BMPs, due to these effects, including a zone of avoidance at the
highest inclinations. Specifically the Kozai evolution could explain
the recently observed peculiar orbit of the Kuiper belt binary 2001
QW$_{322}$ . Additionally, the KCTF process could lead to BMPs coalescence
and serve as an important route for the formation of irregular shaped
single minor planets with large axial tilts. 
\end{abstract}

\section{Introduction}

Stable gravitational triple systems require a hierarchical configuration,
in which two objects orbit each other in a relatively tight {}``inner
binary'', and the third object orbits the binary in a wider {}``outer
binary''. Although such triples are stable against disruption, their
orbits may change shape and orientation on time scales much longer
than their orbital period. In particular, the Kozai-Lidov mechanism\citep{1962K,1962L}
predicts a secular perturbations of the inner binary orbit. So-called
{}``Kozai oscillations'' cause the eccentricity and the inclination
to fluctuate. The Kozai mechanism is known to be highly important
in the evolution of many triple systems\citep{1962K,1979MS,1968H,1998KEM,car+02,nes+03,2007FT}.
It leads to a large (order unity) periodic oscillations (Kozai cycles)
in the eccentricity and inclination of inner binaries with high inclinations
with respect to the outer binary orbit (hereafter the relative inclination).

In recent years many binary minor planets {[}BMPs; both binary asteroids
and binary trans-Neptunian objects (TNOs){]} have been discovered
in the Solar system\citep{ric+06}. BMPs can be regarded as the inner
binary members of a triple system in which the sun is the perturbing
companion in the outer orbit. Here we study the importance of such
perturbations and show that BMPs are susceptible to Kozai oscillations,
which play a major role in their evolution. The combined effects of
the Kozai mechanism in addition to tidal friction, (Kozai cycles and
tidal friction; KCTF\citep{1979MS,1998KEM}) which becomes important
for BMPs with high eccentricities (induced by the Kozai mechanism),
change the orbital parameters of the BMPs. These effects could then
erase the observational signatures suggested by the various formation
scenarios of BMPs discussed in the literature\citep{wei+89,wei02,mer+02,gol+02,fun+04,ric+06,lee+07}.
replacing them with unique and different signatures, that are consistent
with current observations.

\section{Kozai oscillations}

\begin{figure}
\includegraphics[clip,scale=0.4]{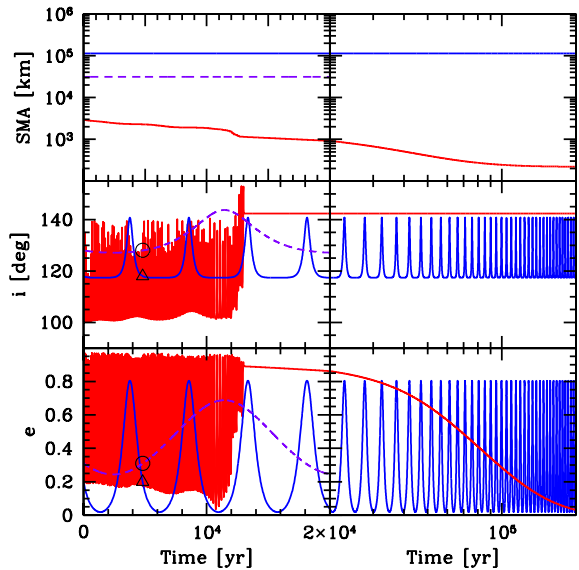}

\caption{\label{f:eccentricity-evolution}Typical Kozai evolution of three
BMPs (note logarithmic time scale on the right). Bold blue curves
show the Kozai cycles of the binary TNO similar to 2001 QW$_{322}$
{[}triangle marks its currently observed\citep{pet+08} orbital parameters{]}.
The evolution of 2001 QW$_{322}$ is not affected by tidal friction,
and probes a wide range of inclinations and eccentricities. The evolution
of the TNO binary 2001 QT$_{297}$ is also shown in dashed line {[}circle
marks its currently observed orbital paramters{]}. Red curves show
the evolution of a hypothetical main belt binary asteroid affected
both by the Kozai mechanism and by tidal friction. The BMP periodically
evolves into high eccentricities and at later times circularize due
to tidal forces and migrates to small SMA separation. Note that the
final inclination is typically close to $140^{\circ}$ at which eccentricities
during the Kozai cycle are highest and tidal friction is therefore
more efficient. The small inclination oscillations observed at the
final stages are due to tidal torques which become important at this
stage\citep{2007FT}. These could have even larger amplitudes if the
spin rate of the BMP members is larger. }

\end{figure}

Large Kozai oscillations (see fig. \ref{f:eccentricity-evolution};
for which we used the Kozai and KCTF evolution code\citep{2007FT})
take place when the relative inclination between the inner binary
orbit and the outer binary orbit of a triple system is large for initially
circular binaries, i.e. $40^{\circ}\lesssim i\lesssim140^{\circ}$
(hereafter Kozai inclinations), with somewhat wider inclination range
for initially eccentric BMPs. Systems with Kozai inclinations keep
their semi-major axis (SMA) separation, but could be driven into periodic
changes of the other orbital parameters of the inner binary. The eccentricity
and inclination change in a specific range which depends on these
orbital parameters. The maximal eccentricity induced by the Kozai
mechanism (assuming small initial eccentricity) is given by \begin{equation}
e_{max}=\sqrt{1-(5/3)cos^{2}(i_{0})},\label{eq:emax}\end{equation}
 where $e_{0}$ and $i_{0}$ are the initial eccentricity and inclination,
respectively (the more general expressions for the maximal eccentricity
and minimal eccentricity (and inclinations) for arbitrary initial
paramters have somewhat more lengthy analytic formulation \citealp{per+08}).
The typical timescale for Kozai oscillations between the limiting
values is given by\citep{1998KEM} \begin{equation}
P_{K}=\frac{2P_{out}^{2}}{3\pi P_{in}}\frac{m_{1}+m_{2}+m_{3}}{m_{3}}(1-e_{out}^{2})^{3/2}\ ,\label{eq:Pk}\end{equation}
 where $P_{out}$ and $P_{in}$ are the orbital periods of the outer
and inner binaries in the triple system, respectively; $m_{1}+\ m_{2}$
and $m_{3}$ are the masses of the inner binary and the third outer
member of the triple (the sun in the case of the BMP-sun triple system
considered here), respectively; and $e_{out}$ is the eccentricity
of the outer binary.

Many of the BMPs observed in the Solar system are known to have large
relative inclinations, making them susceptible to Kozai oscillations
(see \citealp{per+08} for an extended discussion of the observed
inclinations of BMPs). For such systems the timescale for periodic
changes in the absence of any additional forces (such as tidal friction
or pericenter precession for non-spherical objects), could be as short
as a few thousand years (typical for main belt binary asteroids) and
up to $10^{7}$ years for some binary TNOs; much shorter than the
lifetime of these systems (see \ref{f:eccentricity-evolution} for
example). Note that the short, $\sim46$ yrs, Kozai timescale for
the BMP Huenna %
\footnote{The inclination of Huenna is outside the Kozai region for initially
circular binaries, however eccentric binaries are affected by the
Kozai mechanism even at lower eccentricities.%
} may enable a direct observation of the Kozai effect within the next
few years, even with current observations accuracy (we find $e_{\mx}\simeq0.23$
and $e_{\mn}\simeq0.15$ for Huenna; i.e. eccentricity change rate
of $\sim3.4\times10^{-3}$ yr$^{-1}$, where current error bars on
the eccentricity are of $\pm6\times10^{-3}$; \citealp{mar+08}).
The change in eccentricities and inclinations during these timescales
could be as large as $\Delta{e}=e_{\mx}-e_{\mn}\simeq0.99-0.05=0.94$
and $\Delta{i}=i_{\mx}-i_{\mn}\simeq86-39=47{}^{\circ}$ for the known
systems (e.g. for the binary TNO 2000 OJ67; based upon the orbital
parameters determined by \citealp{gru+09}).

\section{Kozai cycles and tidal friction}

The SMA separation of BMPs could evolve due to the Kozai mechanism,
when tidal friction effects are accounted for. Such effects become
important when the pericenter distance between the BMP members becomes
small enough during the Kozai cycles. Orbital energy is then dissipated
in each pericenter approach by the tidal friction, leading to the
evolution of the BMP to smaller SMA separations and more circular
orbits (hereafter Kozai migration or KCTF mechanism\citep{1979MS,1998KEM}
; see fig. \ref{f:eccentricity-evolution}). Moreover, such evolution
could even lead to mass exchange or coalescence of the BMP members
into a single (probably irregularly shaped) minor planet if the separation
becomes small enough.

\begin{figure}
\includegraphics[clip,scale=0.4]{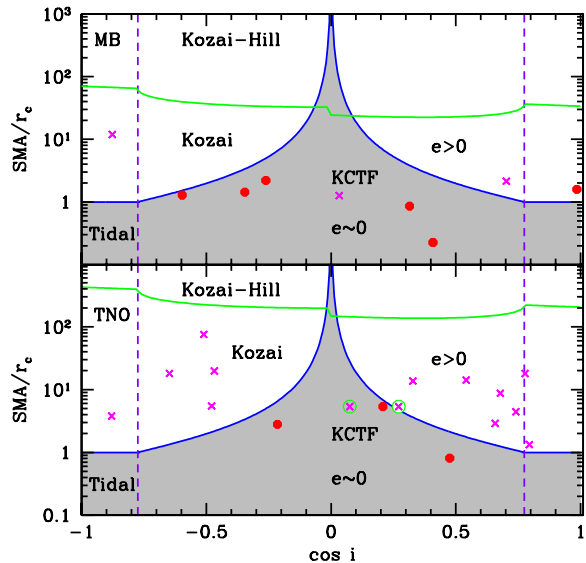}

\caption{\label{f:inclination-separation} The inclination and semi-major axis
phase space at which the Kozai, KCTF and Kozai-Hill mechanisms have
a dominant role for two observed BMP populations; the binary TNOs
(bottom) and the main belt binary asteroids (top). Observed eccentric
( $\times$ ) and circular (filled circles) BMP orbits are shown.
BMPs with two inclinations solutions, in which different solutions
fall in different regimes are marked by green empty circles (otherwise
only one solution is shown, corresponding to better $\chi^{2}$ score;
currently observed BMPs will be discussed extensively in a forthcoming
paper; Naoz \& Perets, in prep. ). BMPs outside the Kozai region (Kozai
inclinations; marked by dashed lines between $40^{\circ}\lesssim i\lesssim140^{\circ}$)
are negligibly affected by the Kozai evolution. BMPs in the Kozai
region can fundamentally change their inclinations and eccentricities
and typically have large eccentricities ($>0.05)$. BMPs in the shaded
region are affected by the Kozai mechanism and/or tidal friction and
are expected to have relatively small SMA separations and circularized
orbits $(<0.05)$. Stable BMPs should not exist in the the Kozai-Hill
region (above the upper arc). The SMA is given in units of the KCTF
region baseline, $r_{c}$ at which BMPs are expected to be circularized. }

\end{figure}

In fig. \ref{f:inclination-separation} we show the regions in orbital
phase space where the Kozai and KCTF processes become important (only
dependence on SMAs and inclinations is shown, eccentricities are assumed
to initially be negligible for simplicity). BMPs in the KCTF region,
i.e. BMPs with large inclinations and small SMAs such that during
their Kozai evolution their pericenter distance becomes small enough
(when their eccentricities become large) for tidal friction to dominate
their further evolution (see fig. \ref{f:inclination-separation})
are expected to have small SMA separations and circularized orbits
(i.e. low eccentricities). Therefore the KCTF regime is expected to
have a zone of avoidance in it's upper region (large SMA separations)
from which BMPs are expected to migrate into its lower region (small
separations), which is expected to be overpopulated with close BMPs
(possibly even contact or coalesced configurations), with low eccentricities.
Moreover, the distribution of the inclinations for migrating BMPs
in the KCTF could be double peaked, increasing towards $40^{\circ}$
and $140^{\circ}$\citep{2007FT} , where the pericenter distance
is smallest during the Kozai cycle, and therefore tidal friction becomes
more dominant and may lock the system from further evolution of the
inclination at the final stages of the KCTF evolution (see fig. \ref{f:eccentricity-evolution}
for such behavior). Note, however, that for fast spinning member of
BMPs and/or low mass ratio BMPs, inclinations may continue to fluctuate
even at late times of the KCTF evolution, and the inclinations distribution
could be strongly affected. In addition, the Kozai mechanism may still
excite small eccentricities in BMPs in the KCTF regime. Such excitations
could be the source of small but non-zero eccentricities of BMPs in
the KCTF regime, such as observed for the binary asteroid Emma and
the the Pluto-Charon system in the Kuiper belt (\citealp{ala+03,mar+08}).
Such excitations are likely to be higher for BMPs with low mass ratios. 

We note that the (usually unknown) asphericity of the binaries components
may also affect the BMPs orbital evolution \citep{cha+93,sch+94,rag+09}.
In addition, minor planets in higher multiplicities (e.g. the triple
TNO systems of 2003 EL61) are subjected to mutual forces from additional
satellites, not taken into account in our description, which can substantially
affect the evolution of the system \citep{rag+09}. Such effects are
beyond the scope of this paper, where we assume all systems to be
binaries with spherical components.

\section{Kozai-Hill instability}

At the other end, if the apocenter distance of a BMP becomes large
enough during the Kozai cycles, the tidal effects of the sun may become
dominant. The binary separation then becomes closer to or even larger
than the Hill radius of the system; and the BMP is disrupted (hereafter
the Kozai-Hill scenario; Perets and Naoz, in preparation). Therefore
the Kozai-Hill region (see fig. \ref{f:inclination-separation}) should
be depleted of BMPs. Note that the Hill radius has an inclinations
dependence at all inclinations due to the changing strength of the
Coriolis force which is proportional to $cos(i)$\citep{ina80,ham+91}
. The effect of the Kozai mechanism is an additional effect which
effects only the Kozai region. We use an analytic expression for the
critical SMA at which a given BMP becomes unstable, which takes into
account both these effects \citep{per+08}. We note that currently
observed BMPs indeed have separations that are a few times smaller
than the Hill radius, but this could also be related to other processes,
such as disruption of the widest binaries through gravitational encounters
with other minor or major planets.

\section{Discussion}

\subsection{KCTF observational signature and BMPs formations scenario}

The Kozai, KCTF and Kozai-Hill processes could appreciably affect
the survival of BMPs, their orbits and their orbital phase-space distribution.
In fig.~\ref{f:inclination-separation} we show typical examples
of the inclination-SMA separation phase-space in which the different
Kozai processes take place. The positions of all BMPs with known inclinations
in this phase-space are also shown (detailed discussion of the observed
inclinations of BMPs is given elsewhere \citealp{per+08}). BMPs in
the Kozai region are not susceptible to tidal friction during the
Kozai cycle. For such BMPs the eccentricity and inclinations continuously
and periodically change between the minimal and maximal Kozai eccentricity
(see e.g. fig.~\ref{f:eccentricity-evolution}) and should therefore
have a wide distribution of eccentricities which would be very different
from the initial conditions set by the formation mechanism of the
BMPs. The initial SMA, however, will be conserved in this case, still
reflecting the initial conditions.

We note that the current observations (although currently only a small
sample exists) are in good agreement with our predictions. All the
BMPs in the Kozai region have relatively high eccentricities (none
are circular, defined here as $e<0.05$). The zone of avoidance at
the highest inclinations in the KCTF region is empty, and BMPs in
the KCTF region tend to be circular and with smaller SMA separations.
The large difference between binary TNOs and binary asteroids is likely
to be related to different formation mechanisms\citep{ric+06} . Note
that BMPs formed only at small separations at which tidal friction
is important would never be subjected to the pure Kozai mechanism.

The above predictions should be of major concern when trying to constrain
BMPs' formation scenarios by observations. Many formations scenarios
for BMPs have been described in the literature (see ref. \citep{ric+06}
for a review) with different predictions for the distribution of BMP
SMA separations, eccentricities and inclinations (unfortunately, only
few studies explored inclination distribution\citep{naz+07,sch+08}
; given the importance of the Kozai effect we strongly suggest to
explore this in future studies). As we have shown the initial conditions
as prescribed by different scenarios will be fundamentally changed
by the Kozai and KCTF evolution. In order to find evidence for specific
formation scenarios of BMPs, it is therefore essential to take the
Kozai evolution of the BMPs into account. For example, typical high
eccentricity BMPs produced in some scenarios such as the expected
$e>0.8$ in the exchange scenario of Funato et al.\citep{fun+04}
, or $0.2\lesssim e\lesssim0.8$ in the chaos assisted capture scenario\citep{lee+07}
, could be observed with very low eccentricities, either due to the
random phase in which they are observed during their Kozai cycle or
due to their KCTF evolution which would have circularized their orbit
(e.g. fig. \ref{f:inclination-separation}). Recently (and after the
initial presentation of this manuscript), \citet{pet+08} have reported
on the peculiar orbit of the Kuiper Belt Binary 2001 QW$_{322}$,
with very large SMA but relatively small eccentricity. Such orbit
is difficult to explain through previously suggested BMPs formation
scenarios, since a binary with such wide SMA is likely to be produced
through the exchange scenario scenario, which would typically produce
a highly eccentric binary \citep{fun+04}. Such orbit, however, is
naturally produced through the Kozai mechanism presented here. As
shown in fig. \ref{f:eccentricity-evolution}, the orbit of a binary
similar to 2001 QW$_{322}$ could evolve due to the Kozai mechanism
from initially high eccentricity and small pericenter distance separation
(in fact comparable to the separation of the \textquotedblleft{}regular\textquotedblright{}
Kuiper belt object 2001 QT$_{297}$) to have a low eccentricity and
much larger pericenter distance separation, as observed today. Such
a binary could therefore have been formed through a binary-single
encounter, and then naturally evolve to its current state through
Kozai evolution. 

Similarly, typical low eccentricity BMPs, such as might be expected
to be produced in the dynamical friction capture scenario\citep{gol+02}
, could be observed with very high eccentricities. 
Since these effects occur mainly for BMPs in the main Kozai region,
we predict an inclination and SMA separation dependence of the eccentricity
distribution of BMPs, with a transition between BMPs at the different
regimes described in fig.~\ref{f:inclination-separation}. Similarly
the SMA separation distribution would also be changed between the
different regimes. In fact, only BMPs outside the Kozai region would
preserve the eccentricity and SMA separation signature of their formation
scenario, whereas in the other regimes these distributions would be
dominated by the Kozai mechanism.

In addition to these predictions, we note the importance of environmental
effects when combined with Kozai evolution. In some regions of the
solar system BMPs were likely to form and/or to evolve for some time
in dense environments where they encountered other minor planets (as
suggested or required by most BMP formation and evolutionary scenarios).
In such an environment, encounters may change the orbital configuration
of the binaries in quite a chaotic way\citep{heg75} especially for
the wider binaries that are more susceptible to encounters (larger
cross section). The Kozai timescales (Eq.~\ref{eq:Pk}) could be
shorter than the typical timescale between consequent encounters.
In such cases a BMP entering the KCTF regime for the first time following
an encounter, could rapidly migrate to smaller SMA separation, thus
decreasing its chance for further encounters which could have otherwise
potentially export it outside this phase space region. This KCTF region
therefore produces a sink in the phase-space distribution and induces
a flow from small inclinations to high inclinations in the phase-space
distribution of BMPs. We predict that the KCTF mechanism, when at
work in a collisional environment, would form a KCTF-collisional signature
in which the KCTF and the phase-space regions close to it (i.e. mostly
the Kozai region) would be overpopulated relative to the other phase
space regions (beside the zone of avoidance which should be depleted
of BMPs). The distribution of observed binary TNOs is suggestive of
such a signature (fig. \ref{f:inclination-separation}), with all
BMPs at high inclinations, in addition to an empty region at the zone
of avoidance, but future observations are required in order to show
this at high significance.

\subsection{Binary spin-orbit correlation and irregularly shaped minor planets }

Since KCTF evolution drives BMPs into close configurations, they may
become close enough as to evolve into contact configuration or even
mergers (similar suggestion was discussed in the context of merger
of inner binary in triple stellar systems due to KCTF; \citealp{per+09}).
These contact/merger products may still show a signature of their
KCTF formation scenario producing a relative spin-orbit inclination
which is likely to be limited to Kozai inclinations. In addition this
Kozai-induced merger scenario could naturally explain the existence
of large single minor planets with irregular prolonged shapes (e.g.
\citep{har+80,rom+01} ) and predict their prolonged axis to be aligned
perpendicular to their spin axis (where the latter is expected to
have the same biased spin-orbit inclination distribution as the Kozai
contact binaries). We note that rotational disruption and direct collisions
could also produce irregularly shaped minor planets, but are not expected
to produce a bias towards high latitude inclinations. The YORP effect
could also form a bias toward specific high inclinations, however
this process is efficient only for small asteroids (diameter smaller
than $50$~km; \citealp[e.g. ][and references there in]{pol+09})
, and is not expected to affect TNOs at all. If such mergers products
are abundant they could effect the relative spin-orbit inclination
distribution, and produce tilt axis distribution which is biased towards
high latitudes from the orbital plane. Interestingly, such a bias
exists in the pole distribution of large binary asteroids \citep[; not expected to be affected by YORP]{kry+07}.

\section{Summary}

In this letter we have shown that the secular Kozai perturbations
by the sun have a major role in the dynamical evolution of binary
minor planets. The Kozai mechanism leads to large periodic oscillations
in the eccentricity and inclination of highly inclined BMP orbits
and the combined effects of the Kozai mechanism and tidal friction
drives BMPs into short period circular orbits. We predict an inclination
dependent distribution of the separation and eccentricity of BMPs,
due to these effects, including a zone of avoidance at the highest
inclinations. Additionally, the KCTF process could lead to BMPs coalescence
and serve as an important route for the formation of irregular shaped
single minor planets with large axial tilts.

\acknowledgements{We would like to thank Dan Fabrycky for supplying 
us with his KCTF evolution code and for his critical review on an early 
version of this manuscript. We also acknowledge valuable discussions 
with Darin Ragozzine, Re'em Sari, Tsevi Mazeh and David Polishook. 
SN acknowledges support from the ISF grant 629/05. We also acknowledge 
the generous support by the industrial and commercial club of Israel 
through the Ilan-Ramon scholarship. }

\bibliography{bss}

\end{document}